\documentclass[12 pt,twoside,aps,prd,amsmath,amssymb,
tightenlines,showpacs,showkeys,eqsecnum]{revtex4-1}
\usepackage{mathptmx}
\usepackage{bm}
\usepackage{graphicx}

\usepackage{hyperref}
\renewcommand{\cal}{\mathcal}

\newcommand {\pr}{\partial}

\newcommand {\cG}{\cal G}
\newcommand {\cL}{\cal L}

\newcommand {\vf}{\varphi}

\newcommand {\vk}{\varkappa}

\def \myfigures #1#2#3#4#5#6#7#8
{\begin{figure}[ht]
    \begin{center}
        \includegraphics[width=#2 \textwidth]{#1.eps}
        \hfill
        \includegraphics[width=#6 \textwidth]{#5.eps}
        \parbox[t]{#4\textwidth}{\caption {#3}\label{#1}}
        \hfill
        \parbox[t]{#8\textwidth}{\caption {#7}\label{#5}}
    \end{center}
\end{figure} }

\def\myfigure #1#2#3#4
{\begin{figure}[ht]\begin{center}
\includegraphics[width=#2 \textwidth]{#1.jpg}
\parbox[t]{#4\textwidth}{\caption{#3}\label{#1}}
\end{center}\end{figure}}

\begin{document}
\baselineskip -24pt
\title{Interacting Scalar and
Electromagnetic Fields in $f(R,\,T)$ Theory of Gravity}
\author{Bijan Saha}
\affiliation{Laboratory of Information Technologies\\
Joint Institute for Nuclear Research, Dubna  \\ 141980 Dubna, Moscow
region, Russia} \email{bijan@jinr.ru}
\homepage{http://bijansaha.narod.ru}

\begin{abstract}
Within the scope of $f(R,\,T)$ gravity we have studied the
interacting scalar and electromagnetic fields in a Bianchi type I
universe. It was found that if the study is confined to the case
$f(R,\,T) = R + \lambda f(T)$, the system is completely given by the
equations similar to Einstein gravity. Moreover, the present study
imposes some severe restrictions on the field equations as well.

\end{abstract}



\maketitle

\bigskip

\section{Introduction}

The existence of dark matter in the Universe as well as the recent
observational data
\cite{Perl1,Perl2,Perl3,Reiss1,Reiss2,Tonry,Clocch} supporting the
accelerated mode of expansion of the Universe pose a fundamental
theoretical challenge to the Einstein theory of gravity. One of the
possible ways to explain the observations is the modification of the
Einstein gravity in such a way that it would give the gravitational
alternative to DE. The modified theories of gravity justify the
unification of DM and DE, transition from deceleration to
acceleration epoch of the universe, description of hierarchy
problem, dominance of effective DE, which help to solve the
coincidence problem and many more. Currently there are a number of
candidates for DE such as cosmological constant, quintessence,
Chaplygin gas, $k$-essence, spinor field, tachyon etc. Another
approach is to modify the general relativity itself. This is known
as modified gravity. One of such theoretical model is known as
$f(R)$ gravity, in which the standard Einstein-Hilbert action is
replaced by an arbitrary function of the Ricci scalar $R$. This
model has been extensively used in recent time and it was found that
the late-time acceleration of the Universe can be explained within
this theory \cite{carrol}. Recently a generalization of $f(R,\,T)$
theory of gravity was proposed by Harko {\it et al} \cite{harko},
where $T$ is the trace of stress-energy tensor. After this paper was
published in 2011, many authors have investigated different problems
within the scope of $f(R,\,T)$ theory. The purpose of this note is
to study the system of interacting scalar and electromagnetic field
within the framework of this new theory and see if this new model
can improve the results obtained earlier for a standard
Einstein-Hilbert model \cite{SahaCEJP2011}.

\section{Basic equations}

Following \cite{harko} let us consider the action of the form

\begin{equation}
S = \frac{1}{2\vk} \int f(R,\,T) \sqrt{-g} d^4 x + \int L_{\rm m}
\sqrt{-g} d^4 x, \label{action}
\end{equation}
with $R$ being the Ricci scalar curvature and $T = g^{\mu\nu}
T_{\mu\nu}$, where the energy-momentum tensor $T_{\mu\nu}$ is given
by

\begin{equation}
T_{\mu\nu} = -\frac{2}{ \sqrt{-g}} \frac{\delta (\sqrt{-g} L_{\rm
m})}{\delta g^{\mu \nu}} = L_{\rm m} g_{\mu \nu} - 2 \frac{\partial
L_{\rm m}}{\partial g^{\mu\nu}}. \label{temmat}
\end{equation}

It should be noted that the basic equations given here are almost
the same as in \cite{harko}, though at places there are some small
differences. That is why I will derive the equations in details and
underline the differences in due course.

Variation of action \eqref{action} with respect to $g^{\mu \nu}$
gives
\begin{equation}
\delta S = \frac{1}{2\vk} \int \left( f_R \delta R + f_T
\frac{\delta T}{\delta g^{\mu \nu}} \delta g^{\mu \nu} +
\frac{1}{\sqrt{-g}} f \delta \sqrt{-g} +
\frac{2\vk}{\sqrt{-g}}\frac{\delta (\sqrt{-g} L_{\rm m})}{\delta
g^{\mu \nu}}\right)\sqrt{-g} d^4 x. \label{varac}
\end{equation}

Further taking into account that $R = g^{\mu \nu} R_{\mu \nu}$ and
$\nabla_\rho g^{\mu \nu} = 0$ one finds that
\begin{equation}
\delta R = \left( R_{\mu \nu} + g_{\mu \nu} \Box - \nabla_\mu
\nabla_\nu\right) \delta g^{\mu \nu}, \label{delr}
\end{equation}
where $\Box = g^{\mu\nu} \nabla_{\mu} \nabla_{\nu}.$

On the other hand
\begin{equation}
\frac{\delta T}{\delta g^{\mu \nu}} = \frac{\delta (g^{\alpha\beta}
T_{\alpha\beta})}{\delta g^{\mu \nu}} = \frac{\delta
g^{\alpha\beta}}{\delta g^{\mu \nu}}  T_{\alpha\beta} +
g^{\alpha\beta} \frac{\delta
 T_{\alpha\beta}}{\delta g^{\mu \nu}} = T_{\mu\nu} +
\Theta_{\mu\nu}, \label{delT}
\end{equation}
where we define
\begin{equation}
\Theta_{\mu\nu} = g^{\alpha\beta} \frac{\delta
T_{\alpha\beta}}{\delta g^{\mu \nu}}, \label{Theta}
\end{equation}

Here let us make a remark on $\delta g^{\alpha\beta}/\delta g^{\mu
\nu}$. Here we define
\begin{equation}
\frac{\delta g^{\alpha\beta}}{\delta g^{\mu \nu}} =
\delta^\alpha_\mu \delta^\beta_\nu. \label{Kron01}
\end{equation}
Like this case we come to the expression
\begin{equation}
\frac{\delta g^{\alpha\beta}}{\delta g^{\mu \nu}}  T_{\alpha\beta} =
T_{\mu\nu}, \label{delT1}
\end{equation}
Alternatively we can define
\begin{equation}
\frac{\delta g^{\alpha\beta}}{\delta g^{\mu \nu}} =
\frac{1}{2}\left[ \delta^\alpha_\mu \delta^\beta_\nu +
 \delta^\alpha_\nu \delta^\beta_\mu\right],
\label{Kron02}
\end{equation}
which will leave the results unaltered. In \cite{harko}, the authors
name $\frac{\delta g^{\alpha\beta}}{\delta g^{\mu \nu}} =
\delta^{\alpha \beta}_{\mu\nu}$ the Generalized Kronecker Symbol,
which in fact is totally antisymmetric (see Mathematical
encyclopedia)
\begin{equation}
\frac{\delta g^{\alpha\beta}}{\delta g^{\mu \nu}} = \delta^{\alpha
\beta}_{\mu\nu} = \left[ \delta^\alpha_\mu \delta^\beta_\nu -
 \delta^\alpha_\nu \delta^\beta_\mu\right],
\label{Kron03}
\end{equation}
and in no way can give the results they have. In this case for
example we have $\frac{\delta g^{\alpha\beta}}{\delta g^{\mu \nu}}
T_{\alpha\beta} = 0$.

Hence after the integration we get the following equation

\begin{eqnarray}
f_R (R,\,T) R_{\mu \nu} &-& \frac{1}{2} f(R,\,T) g_{\mu \nu} +
(g_{\mu
\nu} \Box - \nabla_\mu \nabla_\nu) f_R (R,\,T)\nonumber \\
 &=& \vk
T_{\mu\nu} - f_T (R,\,T) (T_{\mu\nu} + \Theta_{\mu\nu}) \label{FRT1}
\end{eqnarray}

Introducing $G_{\mu \nu} = R_{\mu \nu} - \frac{1}{2} g_{\mu \nu} R$
this equation can be rewritten as

\begin{eqnarray}
f_R (R,\,T) G_{\mu \nu} &-& \frac{1}{2} \left(f(R,\,T) - R f_R
(R,\,T) \right) g_{\mu \nu} + \left(g_{\mu
\nu} \Box - \nabla_\mu \nabla_\nu\right) f_R (R,\,T)\nonumber \\
 &=& \vk
T_{\mu\nu} - f_T (R,\,T) (T_{\mu\nu} + \Theta_{\mu\nu})
\label{FRTEE}
\end{eqnarray}

Taking the trace of \eqref{FRT1} one finds

\begin{eqnarray}
\Box f_R (R,\,T) = \frac{2}{3} f (R,\,T) - \frac{1}{3} f_R (R,\,T)R
+ \frac{\vk}{3} T  - \frac{1}{3} f_T (R,\,T) T - \frac{1}{3} f_T
(R,\,T) \Theta. \label{FRT1trace}
\end{eqnarray}

Then inserting \eqref{FRT1trace} into \eqref{FRT1} one finds

\begin{eqnarray}
f_R (R,\,T) G_{\mu \nu} &+& \frac{g_{\mu \nu}}{6} \left[f(R,\,T) +
f_R(R,\,T) R\right] -  \nabla_\mu \nabla_\nu f_R (R,\,T)\nonumber \\
 &=& \left(\vk - f_T (R,\,T)\right)\left[
T_{\mu\nu} - \frac{1}{3} g_{\mu\nu} T\right] - f_T (R,\,T)\left[
\Theta_{\mu\nu}- \frac{1}{3} g_{\mu\nu} \Theta\right]. \label{FRT2}
\end{eqnarray}

It should be noted that the Eq. \eqref{FRT2} can be used only in
case $f_R (R,\,T) \ne {\rm const.}$, otherwise, i.e., if $f_R
(R,\,T) = {\rm const.}$ the corresponding equation will be

\begin{eqnarray}
f_R (R,\,T) R_{\mu \nu} - \frac{1}{2} f(R,\,T) g_{\mu \nu}  = \vk
T_{\mu\nu} - f_T (R,\,T) (T_{\mu\nu} + \Theta_{\mu\nu})
\label{FRT1n}
\end{eqnarray}

And finally, taking into account that

$$\nabla_\mu \nabla_\nu f_R(R,T) = \frac{\partial^2 f_R (R,T)}{\partial x^\mu \partial x^\nu}
+ \Gamma_{\mu \nu}^{\alpha}\frac{\partial f_R (R,T)}{\partial
x^\alpha}$$

we rewrite \eqref{FRT2} in the form

\begin{eqnarray}
G_{\mu \nu} &=& \frac{1}{f_R (R,\,T)} \left[ \frac{\partial^2 f_R
(R,T)}{\partial x^\mu \partial x^\nu} + \Gamma_{\mu
\nu}^{\alpha}\frac{\partial f_R (R,T)}{\partial
x^\alpha}\right] - \frac{g_{\mu \nu}}{6f_R (R,\,T)} \left[f(R,\,T) + f_R(R,\,T) R\right] \nonumber \\
 &+& \frac{\vk - f_T (R,\,T)}{f_R (R,\,T)}\left[
T_{\mu\nu} - \frac{1}{3} g_{\mu\nu} T\right] - \frac{f_T
(R,\,T)}{f_R (R,\,T)}\left[ \Theta_{\mu\nu}- \frac{1}{3} g_{\mu\nu}
\Theta\right] \label{FRT3}
\end{eqnarray}
Further taking into account that
\begin{eqnarray}
\frac{\partial f_R (R,T)}{\partial x^\alpha} = \frac{\partial f_R
(R,T)}{\partial R} \frac{d R}{d x^\alpha} + \frac{\partial f_R
(R,T)}{\partial T} \frac{d T}{d x^\alpha} \nonumber
\end{eqnarray}
and
\begin{eqnarray}
 \frac{\partial^2 f_R (R,T)}{\partial x^\mu \partial x^\nu} &=&
\frac{\partial^2 f_R (R,T)}{\partial R^2} \frac{d R}{d x^\mu}\frac{d
R}{d x^\nu} + \frac{\partial^2 f_R (R,T)}{\partial T\partial
R}\frac{d T}{d x^\mu}\frac{d R}{d x^\nu} + \frac{\partial f_R
(R,T)}{\partial
R} \frac{d^2 R}{d x^\mu d x^\nu} \nonumber \\
&+& \frac{\partial^2 f_R (R,T)}{\partial T^2} \frac{d T}{d
x^\mu}\frac{d T}{d x^\nu} + \frac{\partial^2 f_R (R,T)}{\partial R
\partial T} \frac{d R}{d x^\mu}\frac{d T}{d x^\nu} +
\frac{\partial f_R (R,T)}{\partial T} \frac{d^2 T}{d x^\mu d x^\nu}
\nonumber
\end{eqnarray}
we finally obtain
\begin{eqnarray}
G_{\mu \nu} &=& \frac{1}{f_R (R,\,T)} \Biggl[\frac{\partial^2 f_R
(R,T)}{\partial R^2} \frac{d R}{d x^\mu}\frac{d R}{d x^\nu} +
\frac{\partial^2 f_R (R,T)}{\partial T\partial R}\frac{d T}{d
x^\mu}\frac{d R}{d x^\nu} + \frac{\partial f_R (R,T)}{\partial
R} \frac{d^2 R}{d x^\mu d x^\nu} \nonumber \\
&+& \frac{\partial^2 f_R (R,T)}{\partial T^2} \frac{d T}{d
x^\mu}\frac{d T}{d x^\nu} + \frac{\partial^2 f_R (R,T)}{\partial R
\partial T} \frac{d R}{d x^\mu}\frac{d T}{d x^\nu} +
\frac{\partial f_R (R,T)}{\partial
T} \frac{d^2 T}{d x^\mu d x^\nu} \nonumber \\
&+& \Gamma_{\mu \nu}^{\alpha} \left[ \frac{\partial f_R
(R,T)}{\partial R} \frac{d R}{d x^\alpha} + \frac{\partial f_R
(R,T)}{\partial T} \frac{d T}{d x^\alpha} \right]  \Biggr]
- \frac{ g_{\mu \nu}}{6f_R (R,\,T)}\left[f(R,\,T) + f_R(R,\,T) R\right] \nonumber \\
 &+& \frac{\vk - f_T (R,\,T)}{f_R (R,\,T)}\left[
T_{\mu\nu} - \frac{1}{3} g_{\mu\nu} T\right] - \frac{f_T
(R,\,T)}{f_R (R,\,T)}\left[ \Theta_{\mu\nu}- \frac{1}{3} g_{\mu\nu}
\Theta\right] \label{FRT4}
\end{eqnarray}

Divergence of \eqref{FRTEE} leads to

\begin{eqnarray}
\left(\nabla^\mu f_R (R,\,T)\right) G_{\mu \nu} &-& \frac{1}{2}
\left(\nabla_\nu f(R,\,T) - (\nabla_\nu R) f_R (R,T) - R  \nabla_\nu
f_R (R,\, T)\right) +
\left(\nabla_\nu \Box - \Box \nabla_\nu\right) f_R (R,\,T)\nonumber \\
 &=& \left(\vk - f_T (R,T)\right)
\nabla^\mu T_{\mu\nu} - \left(\nabla^\mu f_T (R,\,T)\right)
\left(T_{\mu\nu} + \Theta_{\mu\nu}\right)\nonumber \\ & -& f_T
(R,T)\nabla^\mu \Theta_{\mu\nu}, \label{FRT2o}
\end{eqnarray}
where we used the fact that $g_{\mu \nu} \nabla^\mu f (R,\,T) =
\nabla_\nu f (R,\,T) = f_R \nabla_\nu R + f_T \nabla_\nu T = f_R
g_{\mu \nu} \nabla^\mu R + f_T \nabla_\nu T$, \quad $ \nabla^\mu
G_{\mu \nu} = 0$, \quad $ \nabla^\mu g_{\mu \nu} = 0$ and
\begin{eqnarray}
\left(\nabla_\nu \Box - \Box \nabla_\nu\right) f_R (R,\,T) = -
(\nabla^\mu f_R (R,\,T)) R_{\mu\nu}. \label{Boxrel}
\end{eqnarray}
The relation \eqref{Boxrel} follows from the fact that
\begin{eqnarray}
\left(\nabla_\nu \Box - \Box \nabla_\nu\right) S &=& g^{\alpha
\beta}\left(\nabla_\nu \nabla_\alpha \nabla_\beta - \nabla_\alpha
\nabla_\beta \nabla_\nu\right) S \nonumber \\
&=&  g^{\alpha \beta}\left(\nabla_\nu \nabla_\alpha \nabla_\beta -
\nabla_\alpha \nabla_\nu \nabla_\beta\right) S = g^{\alpha
\beta}\left(\nabla_\nu \nabla_\alpha - \nabla_\alpha \nabla_\nu \right) S_{;\beta} \nonumber \\
&=& g^{\alpha \beta} R^\eta_{\beta \alpha \nu} S_{;\eta} =
-R^\eta_\nu S_{;\eta}, \label{Boxrel1}
\end{eqnarray}
where we used the fact that for any scalar $S$
\begin{eqnarray}
\nabla_\beta \nabla_\nu S = \nabla_\nu \nabla_\beta S,
\label{CovScal}
\end{eqnarray}
and for any vector $S_{;\beta}$
\begin{eqnarray}
\nabla_\nu \nabla_\alpha S_{;\beta} - \nabla_\alpha \nabla_\nu
S_{;\beta} = R^\eta_{\beta \alpha \nu} S_{;\eta} = - R^\eta_{\beta
\nu \alpha} S_{;\eta}. \label{CovVec}
\end{eqnarray}
We also denote $\nabla_\beta S = S_{;\beta}$. After a little
manipulation from \eqref{FRT2o} we find

\begin{eqnarray}
\left(\vk - f_T (R,T)\right) \nabla^\mu T_{\mu\nu} =
\left(\nabla^\mu f_T (R,\,T)\right) \left(T_{\mu\nu} +
\Theta_{\mu\nu}\right) +  f_T (R,T)\nabla^\mu \Theta_{\mu\nu} -
\frac{1}{2} f_T \nabla_\nu T, \label{FRT3o}
\end{eqnarray}
which in view of $\nabla^\mu f_T (R,\,T) = f_T (R,\,T) \nabla^\mu
\ln{f_T (R,\,T)} $ gives

\begin{eqnarray}
\nabla^\mu T_{\mu\nu} = \frac{f_T (R,\,T)}{\left(\vk - f_T
(R,\,T)\right)} \left[\left(T_{\mu\nu} + \Theta_{\mu\nu}\right)
\nabla^\mu \ln{f_T (R,\,T)} + \Theta_{\mu\nu} - \frac{1}{2}
\nabla_\nu T\right]. \label{FRT4o}
\end{eqnarray}

Note that unlike many authors here we have an additional term $ -
\frac{1}{2} \nabla_\nu T$ which comes from $ \nabla_\nu f(R,\,T)$.

Let us now calculate the tensor $\Theta_{\mu\nu}$. Varying
\eqref{temmat} with respect to metric function we find
\begin{eqnarray}
\frac{\delta T_{\alpha \beta}}{\delta g^{\mu \nu}} &=&  \frac{\delta
g_{\alpha \beta}}{\delta g^{\mu \nu}} L_{\rm m} + g_{\alpha \beta}
\frac{\partial L_{\rm m}}{\partial g^{\mu \nu}} - 2 \frac{\partial^2
L_{\rm m}}{\partial g^{\alpha \beta}\partial g^{\mu \nu}} \nonumber
\\
&=&\frac{\delta g_{\alpha \beta}}{\delta g^{\mu \nu}} L_{\rm m} +
\frac{1}{2} g_{\alpha \beta} [ g_{\mu \nu} L_{\rm m} - T_{\mu \nu}]
- 2 \frac{\partial^2 L_{\rm m}}{\partial g^{\alpha \beta}\partial
g^{\mu \nu}}. \label{th1}
\end{eqnarray}
From the condition $g_{\alpha \sigma} g^{\sigma \beta} =
\delta_{\alpha}^{\beta}$, we have
\begin{equation}
\frac{\delta g_{\alpha \beta}}{\delta g^{\mu \nu}} = - g_{\alpha
\eta} g_{\beta \tau} \frac{\delta g^{\eta \tau}}{\delta g^{\mu
\nu}}.\label{conco}
\end{equation}
Further using the definition \eqref{Kron01}, we find
\begin{eqnarray}
\Theta_{\mu\nu} = g^{\alpha \beta}\frac{\delta T_{\alpha
\beta}}{\delta g^{\mu \nu}} = - 2 T_{\mu \nu}  + g_{\mu\nu} L_{\rm
m} - 2 g^{\alpha \beta}  \frac{\partial^2 L_{\rm m}}{\partial g^{\mu
\nu}\partial g^{\alpha \beta}}. \label{th2}
\end{eqnarray}
 Further on account of $g_{\mu\nu} L_{\rm m} = T_{\mu\nu} + 2
\frac{\partial L_{\rm m}}{\partial g^{\mu \nu}}$ we write
\begin{eqnarray}
\Theta_{\mu\nu} = - T_{\mu \nu}  + 2\left[ \frac{\partial L_{\rm
m}}{\partial g^{\mu \nu}} -  g^{\alpha \beta} \frac{\partial^2
L_{\rm m}}{\partial g^{\mu \nu}\partial g^{\alpha \beta}}\right].
\label{th3}
\end{eqnarray}

\section{Matter field Lagrangian}

In what follows we consider the electromagnetic field with induced
massive term given by the Lagrangian \cite{SahaIJTP1997}

\begin{equation}
{\cL} =\left( - \frac{1}{4} F_{\eta\tau}F^{\eta\tau} + \frac{1}{2}
\vf_{,\eta} \vf^{,\eta}{\cG}\right), \quad {\cG}= (1 + \Phi( I)),
\quad I = A_\mu A^\mu.\ \label{lagemsc}
\end{equation}
It should be noted that since the early days of elementary particle
physics, attempts were undertaken to construct a divergence-free
theory. A nonlinear modification of Maxwell theory was proposed by
Mie\cite{mie1,mie2}, with the nonlinear electric current of the form
$j_\mu = (A_\nu A^\nu)^2 A_\mu$. Here $\Phi(I)$ is the function of
invariant $I = A_\mu A^\mu$ characterizes the interaction between
the scalar $\vf$ and electromagnetic $A_\mu$ fields. The inclusion
of Mie invariants breaks the gauge symmetry at small distances, the
Maxwell theory being restored at large distances, i,e, in the linear
limit \cite{SahaCEJP2011}.

In this case we find
\begin{eqnarray}
\frac{\partial {\cL}}{\partial g^{\mu \nu}} = \frac{1}{2}\left[-
F_{\mu \eta} F_{\nu}^{\eta} + {\cG} \vf_{,\mu} \vf_{,\nu} +
\vf_{,\eta} \vf^{,\eta}{\cG}_I A_\mu A_\nu\right], \quad {\cG}_I =
\frac{\partial {\cG}}{\partial I}, \label{1}
\end{eqnarray}
and
\begin{eqnarray}
g^{\alpha\beta}\frac{\partial^2 {\cL}}{\partial g^{\mu \nu} \partial
g^{\alpha \beta}} = \frac{1}{2}\left[- F_{\alpha \mu} F^{\alpha}_
{\nu} + {\cG}_I \vf_{,\alpha} \vf^{,\alpha} A_\mu A_\nu + I {\cG}_I
\vf_{,\mu} \vf_{,\nu}  +  I {\cG}_{II} \vf_{,\eta} \vf^{,\eta} A_\mu
A_\nu \right]. \label{2}
\end{eqnarray}

Then for $\Theta_{\mu \nu}$ one gets
\begin{eqnarray}
\Theta_{\mu \nu} = - T_{\mu \nu} + \left({\cG} - I
{\cG}_I\right)\vf_{,\mu} \vf_{,\nu} - I {\cG}_{II} \vf_{,\eta}
\vf^{,\eta} A_\mu A_\nu, \label{thetaint}
\end{eqnarray}
with
\begin{eqnarray}
T_{\mu \nu} = \left[F_{\mu \alpha} F_{\nu}^{\alpha} -
\frac{1}{4}g_{\mu \nu} F_{\alpha \beta} F^{\alpha \beta}\right] +
\left[\frac{1}{2} {\cG} g_{\mu \nu} - {\cG}_I A_\mu A_\nu\right]
\vf_{,\alpha} \vf^{,\alpha} - {\cG} \vf_{,\mu} \vf_{,\nu}.
\label{emtint}
\end{eqnarray}

From \eqref{thetaint} and \eqref{emtint} one finds

\begin{equation}
\Theta = - T + \left({\cG} - I {\cG}_I - I^2
{\cG}_{II}\right)\vf_{,\alpha} \vf^{,\alpha} = - I^2
{\cG}_{II}\vf_{,\alpha} \vf^{,\alpha} \label{ThetaTrace}
\end{equation}
and

\begin{equation}
T =  \left({\cG} - I {\cG}_I\right)\vf_{,\alpha} \vf^{,\alpha}
\label{TTrace}
\end{equation}

In literature it is customary to consider a few cases such as

(i) $f(R,T) = R + \lambda f(T)$,

(ii) $f(R,T) = f_1(R) + f_2(T)$,

(iii) $f(R,\,T) = f_1(R) + f_2(T)f_3(T)$.

In this paper we consider only the simplest case setting $f(R,T) = R
+ \lambda f(T)$.

In this case from \eqref{FRT1} we find

\begin{equation}
G_{\mu\nu} = R_{\mu\nu} - \frac{1}{2} g_{\mu\nu} R = \vk T_{\mu\nu}
- \lambda \left(\Theta_{\mu\nu} + T_{\mu\nu} \right) + \frac{1}{2}
g_{\mu\nu} \lambda T. \label{Ein1}
\end{equation}
From this equation one can easily obtain
\begin{equation}
R_{\mu}^{\nu} = \vk \left( T_{\mu}^{\nu} - \frac{1}{2}
\delta_\mu^\nu T\right) - \lambda \left(\Theta_{\mu}^{\nu} +
T_{\mu}^{\nu} \right) +  \frac{1}{2} \lambda \delta_{\mu}^{\nu}
\Theta . \label{Ein2}
\end{equation}

We consider the BI metric in the form
\begin{equation}
ds^2 = e^{2\alpha} dt^2 - e^{2\beta_1} dx^2 - e^{2\beta_2} dy^2 -
e^{2\beta_3} dz^2. \label{BI}
\end{equation}
The metric functions $\alpha, \beta_1, \beta_2, \beta_3$ depend on
$t$ only and obey the coordinate condition
\begin{equation}
\alpha = \beta_1 + \beta_2 + \beta_3. \label{cc}
\end{equation}
Note that the Ricci tensor for the Biacnchi type-I model given in
the form \eqref{BI} has only nonzero diagonal components. So here we
write only the diagonal components of the system of equations
\eqref{Ein2}, while the off-diagonal components we use later in
order to find the relations between metric functions and the
components of the vector potential. On account of \eqref{cc} we find
\begin{subequations}
\label{BID}
\begin{eqnarray}
e^{-2\alpha} \left(\ddot \alpha - \dot \alpha^2 + \dot \beta_1^2 +
\dot \beta_2^2 + + \dot \beta_3^2 \right) &=&  \vk \left( T_{0}^{0}
- \frac{1}{2} T\right) - \lambda \left(\Theta_{0}^{0} + T_{0}^{0}
\right) +  \frac{1}{2} \lambda \Theta ,\label{00}\\
e^{-2\alpha} \ddot \beta_1 &=&  \vk \left( T_{1}^{1} - \frac{1}{2}
T\right) - \lambda \left(\Theta_{1}^{1} + T_{1}^{1}
\right) +  \frac{1}{2} \lambda \Theta,\label{11}\\
e^{-2\alpha} \ddot \beta_2 &=& \vk \left( T_{2}^{2} - \frac{1}{2}
T\right) - \lambda \left(\Theta_{2}^{2} + T_{2}^{2}
\right) +  \frac{1}{2} \lambda \Theta,\label{22}\\
e^{-2\alpha} \ddot \beta_3 &=&  \vk \left( T_{3}^{3} - \frac{1}{2}
T\right) - \lambda \left(\Theta_{3}^{3} + T_{3}^{3} \right) +
\frac{1}{2} \lambda \Theta,\label{33}
\end{eqnarray}
\end{subequations}
where over dot means differentiation with respect to $t$.

Variation of \eqref{lagemsc} with respect to electromagnetic field
gives
\begin{equation}
\frac{1}{\sqrt{-g}} \frac{\pr}{\pr x^\nu} \left(\sqrt{-g}
F^{\mu\nu}\right) - \left(\vf_{,\nu} \vf^{,\nu}\right) {\cG}_I
A^{\mu} = 0, \quad  {\cG}_I = \frac{d {\cG}}{dI}. \label{emf}
\end{equation}
The scalar field equation corresponding to the Lagrangian
\eqref{lagemsc} has the form
\begin{equation}
\frac{1}{\sqrt{-g}} \frac{\pr}{\pr x^\mu} \left(\sqrt{-g} g^{\mu\nu}
\vf_{,\nu} {\cG}\right) = 0. \label{scf}
\end{equation}

We consider the case when the electromagnetic and scalar fields are
the functions of $t$ only.  Taking this in mind we choose the vector
potential in the following way:
\begin{equation}
A_{\mu} = \left(0,\,A_1(t),\,A_2(t),\,A_3(t)\right). \label{vpot}
\end{equation}
In this case the electromagnetic field tensor $F^{\mu \nu}$ has
three non-vanishing components, namely
\begin{equation}
F_{01} = \dot A_1, \quad F_{02} = \dot A_2, \quad F_{03} = \dot A_3.
\label{emtensor}
\end{equation}
On account of \eqref{vpot} and \eqref{emtensor} we now have
\begin{eqnarray}
I &=&  - A_1^2 e^{-2\beta_1} - A_2^2
e^{-2\beta_2} - A_3^2 e^{-2\beta_3}, \label{eminv}\\
F_{\mu\nu}F^{\mu\nu} &=& - 2 e^{-2\alpha} \left(\dot A_1^2
e^{-2\beta_1} + \dot A_2^2  e^{-2\beta_2} + \dot A_3^2e^{-2\beta_3}
\right). \label{fmn}
\end{eqnarray}
Let us now solve the scalar field equation. Taking into account that
$\vf = \vf(t)$, from the scalar field equation one finds
\begin{equation}
\dot \vf = \frac{\vf_0}{{\cG}}, \qquad \Rightarrow
\vf_{,\mu}\vf^{,\mu} = \frac{\vf_0^2}{{\cG}^2} e^{-2\alpha}, \quad
\vf_0 = {\rm const.} \label{sfs}
\end{equation}
On account of \eqref{emtensor} and \eqref{sfs} from \eqref{emf} for
electromagnetic field we find
\begin{subequations}
\label{emf123}
\begin{eqnarray}
\frac{d}{dt}\left(\dot A_1 e^{-2\beta_1}\right) - \vf_0^2 P_I
A_1 e^{-2\beta_1} &=& 0, \label{emf1}\\
\frac{d}{dt}\left(\dot A_2 e^{-2\beta_2}\right) - \vf_0^2 P_I
A_2 e^{-2\beta_2} &=& 0, \label{emf2}\\
\frac{d}{dt}\left(\dot A_3 e^{-2\beta_3}\right) - \vf_0^2 P_I A_3
e^{-2\beta_3}  &=& 0, \label{emf3}
\end{eqnarray}
\end{subequations}
where we set $P(I) = 1/{\cG(I)}$.

The nonzero components of the energy momentum tensor of material
fields. In view of \eqref{sfs} from \eqref{emtint} we find
\begin{subequations}
\label{emtcomp}
\begin{eqnarray}
T_0^0 &=&\left[\frac{\vf_0^2 P}{2}+\frac{1}{2} \left(\dot A_1^2
e^{-2\beta_1} + \dot A_2^2 e^{-2\beta_2} + \dot
A_3^2 e^{-2\beta_3}\right)\right] e^{-2\alpha}, \label{00emt}\\
T_1^1 &=& \left[- \frac{ \vf_0^2 P}{2} +\frac{1}{2} \left(\dot A_1^2
e^{-2\beta_1} - \dot A_2^2 e^{-2\beta_2} - \dot A_3^2
e^{-2\beta_3}\right)
+\vf_0^2 P_I  A_1^2 e^{-2\beta_1}\right] e^{-2\alpha}, \label{11emt}\\
T_2^2 &=& \left[- \frac{\vf_0^2 P}{2} +\frac{1}{2} \left(\dot A_2^2
e^{-2\beta_2} - \dot A_3^2 e^{-2\beta_3} - \dot A_1^2
e^{-2\beta_1}\right)
+\vf_0^2 P_I A_2^2 e^{- 2\beta_2}\right]e^{-2\alpha}, \label{22emt}\\
T_3^3 &=& \left[- \frac{\vf_0^2 P}{2}  +\frac{1}{2} \left(\dot A_3^2
e^{-2\beta_3} - \dot A_1^2 e^{-2\beta_1} - \dot A_2^2
e^{-2\beta_2}\right)
+ \vf_0^2 P_I  A_3^2 e^{- 2\beta_3}\right]e^{-2\alpha}, \label{33emt}\\
T_2^1 &=& \left(\dot A_1 \dot A_2 + \vf_0^2 P_I A_1
A_2\right) e^{-2 \alpha -2\beta_1},\label{12emt}\\
T_3^2 &=& \left(\dot A_2 \dot A_3 + \vf_0^2 P_I A_2
A_3\right) e^{-2 \alpha -2\beta_2},\label{23emt}\\
T_1^3 &=& \left(\dot A_3 \dot A_1 + \vf_0^2 P_I A_3 A_1\right) e^{-2
\alpha -2\beta_3}.\label{31emt}
\end{eqnarray}
\end{subequations}
From \eqref{emtcomp} one also finds
\begin{equation}
T = \left[ - \vf_0^2 P + \vf_0^2 P_I \left(A_1^2 e^{-2\beta_1} +
A_2^2 e^{-2\beta_2} + A_3^2 e^{-2\beta_3}\right)\right] e^{-2\alpha}
= - \vf_0^2 \left[P + I P_I\right]e^{-2\alpha}. \label{T}
\end{equation}

From \eqref{thetaint} we now find

\begin{subequations}
\label{emthcomp}
\begin{eqnarray}
\Theta_0^0 &=& - T_0^0 + \left({\cG} - I {\cG}_I\right) \frac{\vf_0^2}{{\cG}^2}e^{-2\alpha} =
-T_0^0 + \vf_0^2 \left(P + IP_I\right)e^{-2\alpha}, \label{00emth}\\
\Theta_1^1 &=& - T_1^1 + \frac{\vf_0^2}{{\cG}^2} I {\cG}_{II} A_1^2 e^{-2(\alpha + \beta_1)}, \label{11emth}\\
\Theta_2^2 &=& - T_2^2 + \frac{\vf_0^2}{{\cG}^2} I {\cG}_{II} A_2^2 e^{-2(\alpha + \beta_2)}, \label{22emth}\\
\Theta_3^3 &=& - T_3^3 + \frac{\vf_0^2}{{\cG}^2} I {\cG}_{II} A_3^2 e^{-2(\alpha + \beta_3)}, \label{33emth}\\
\Theta_2^1 &=& - T_2^1 + \frac{\vf_0^2}{{\cG}^2} I {\cG}_{II} A_1 A_2 e^{-2(\alpha + \beta_1)}, \label{12emth}\\
\Theta_3^2 &=& - T_3^2 + \frac{\vf_0^2}{{\cG}^2} I {\cG}_{II} A_2 A_3 e^{-2(\alpha + \beta_2)}, \label{23emth}\\
\Theta_1^3 &=& - T_1^3 + \frac{\vf_0^2}{{\cG}^2} I {\cG}_{II} A_3
A_1 e^{-2(\alpha + \beta_3)}. \label{31emth}
\end{eqnarray}
\end{subequations}
From \eqref{emtcomp} one also finds
\begin{equation}
\Theta = -\frac{\vf_0^2}{{\cG}^2} I^2 {\cG}_{II}e^{-2\alpha} =
-\vf_0^2 I^2 \left[\frac{2}{P} P_I^2 - P_{II}\right]e^{-2\alpha}.
\label{Th}
\end{equation}

Taking into account that the BI metric given by \eqref{BI} has only
nonzero diagonal components from \eqref{Ein1} we have the following
constraints:
\begin{subequations}
\label{cons}
\begin{eqnarray}
\vk T_2^1 - \lambda \left(\Theta_2^1 + T_2^1\right) & = & \left[ \vk
\left(\dot A_1 \dot A_2 + \vf_0^2 P_I A_1 A_2 \right) - \lambda
\frac{\vf_0^2}{{\cG}^2} I {\cG}_{II} A_1 A_2 \right] e^{-2(\alpha
+\beta_1)} = 0,\label{12cons}\\
\vk T_3^2 - \lambda \left(\Theta_3^2 + T_3^2\right) & = & \left[ \vk
\left(\dot A_2 \dot A_3 + \vf_0^2 P_I A_2 A_3 \right) - \lambda
\frac{\vf_0^2}{{\cG}^2} I {\cG}_{II} A_2 A_3 \right] e^{-2(\alpha
+ \beta_2)} = 0,\label{23cons}\\
\vk T_1^3 - \lambda \left(\Theta_1^3 + T_1^3\right) & = & \left[ \vk
\left(\dot A_3 \dot A_1 + \vf_0^2 P_I A_3 A_1 \right) - \lambda
\frac{\vf_0^2}{{\cG}^2} I {\cG}_{II} A_3 A_1 \right] e^{-2 (\alpha +
\beta_3)} = 0.\label{31cons}
\end{eqnarray}
\end{subequations}
From \eqref{cons} one finds

\begin{equation}
\frac{\dot A_1 \dot A_2}{A_1 A_2} = \frac{\dot A_2 \dot A_3}{A_2
A_3} = \frac{\dot A_3 \dot A_1}{A_3 A_1} =
\frac{\lambda}{\vk}\frac{\vf_0^2}{{\cG}^2} I {\cG}_{II} - \vf_0^2
P_I = \frac{\lambda}{\vk} \vf_0^2 I \left[\frac{2}{P} P_I^2 -
P_{II}\right]  - \vf_0^2 P_I . \label{rela123}
\end{equation}
From \eqref{rela123} one finds
\begin{equation}
\frac{\dot A_1}{A_1} = \frac{\dot A_2}{A_2} = \frac{\dot A_3}{A_3}
\equiv \frac{\dot A}{A}, \label{rela123a}
\end{equation}
that leads to the following relations between the three components
of vector potential:
\begin{equation}
A_1 = A, \quad A_2 = C_{21}A, \quad A_3 = C_{31}A,
\label{rela123new}
\end{equation}
with $C_{21}$ and $C_{31}$ being constants of integration. In view
of \eqref{rela123a} we find
\begin{eqnarray}
\dot A_1^2 e^{-2\beta_1} + \dot A_2^2 e^{-2\beta_2} + \dot A_3^2
e^{-2\beta_3} & = &\left(A_1^2 e^{-2\beta_1} + A_2^2 e^{-2\beta_2} +
A_3^2 e^{-2\beta_3}\right)\frac{\dot A^2}{A^2} \nonumber \\ &=&
\left( \vf_0^2 P_I - \frac{\lambda}{\vk} I \left[\frac{2}{P} P_I^2 -
P_{II}\right] \right) I, \label{rela123new1}
\end{eqnarray}
where we have used the relation \eqref{rela123}. On account of
\eqref{rela123new1} further we find
\begin{subequations}
\label{emtcompn}
\begin{eqnarray}
T_0^0 &=& \left[\frac{1}{2} \vf_0^2 \left(P + IP_I\right) -
\frac{\lambda}{2\vk} \vf_0^2 I^2 \left[\frac{2}{P} P_I^2 -
P_{II}\right] \right]e^{-2\alpha}, \label{00emtn}\\
T_1^1 &=& - T_0^0 + \frac{\lambda}{\vk} \vf_0^2 I \left[\frac{2}{P}
P_I^2 - P_{II}\right] A_1^2 e^{-2(\alpha + \beta_1)}, \label{11emtn}\\
T_2^2 &=& - T_0^0 + \frac{\lambda}{\vk}\vf_0^2 I \left[\frac{2}{P}
P_I^2 - P_{II}\right] A_2^2 e^{-2(\alpha + \beta_2)}, \label{22emtn}\\
T_3^3 &=& - T_0^0 + \frac{\lambda}{\vk}\vf_0^2 I \left[\frac{2}{P}
P_I^2 - P_{II}\right] A_3^2 e^{-2(\alpha + \beta_3)}. \label{33emtn}
\end{eqnarray}
\end{subequations}

Inserting \eqref{emtcompn}, \eqref{emthcomp}and \eqref{Th} into
\eqref{BID} for the metric functions one finds:
\begin{subequations}
\begin{eqnarray}
\ddot \alpha - \dot \alpha^2 + \dot \beta_1^2 + \dot \beta_2^2 +
\dot \beta_3^2  &=& (\vk - \lambda)\vf_0^2 [P + I P_I] - \lambda
\vf_0^2   I^2 \left[\frac{2}{P}
P_I^2 - P_{II}\right],\label{00new1}\\
\ddot \beta_1 &=&  0, \label{11new1}\\
\ddot \beta_2 &=& 0, \label{22new1}\\
\ddot \beta_3 &=& 0.  \label{33new1}
\end{eqnarray}
\end{subequations}
From \eqref{11new1}, \eqref{22new1} and \eqref{33new1} we find
\begin{equation}
\beta_1 = b_1 t + \beta_{10}, \quad \beta_2 = b_2 t + \beta_{20},
\quad  \beta_3 = b_3 t + \beta_{30}. \label{betas}
\end{equation}
Here $b_i$ and $\beta_{i0}$ are integration constants. It should be
noted that in order to maintain the same scaling along all the axes,
the constants $\beta_{i0}$ should be the same, hence, without losing
generality one can set $\beta_{i0} = 0$. Let us note that the Eqns.
\eqref{11new1} - \eqref{33new1} coincide with those obtained in
\cite{SahaCEJP2011}, while the \eqref{00new1} has the additional
term, namely $\lambda \vf_0^2 P^2 I^2 {\cG}_{II}$. In case of
$\lambda = 0$ we come to the standard Einstein case, which is
obvious. Moreover, for ${\cG}_{II} = 0$, i.e. ${\cG} = C_1 I + C_2$
be a linear function of invariant $I$ we get the analogous result.

Now let us go back to the electromagnetic field equations
\eqref{emf123}, which can be arranged as
\begin{subequations}
\label{emf123drov}
\begin{eqnarray}
\Bigl(\frac{\dot A_1}{A_1}\Bigr)^{\cdot} + \Bigl(\frac{\dot
A_1}{A_1}\Bigr)^2 - 2 \Bigl(\frac{\dot A_1}{A_1}\Bigr)\dot \beta_1 -
\vf_0^2 P_I &=& 0, \label{emf1drov}\\
\Bigl(\frac{\dot A_2}{A_2}\Bigr)^{\cdot} + \Bigl(\frac{\dot
A_2}{A_2}\Bigr)^2 - 2 \Bigl(\frac{\dot A_2}{A_2}\Bigr)\dot \beta_2 -
\vf_0^2 P_I &=& 0, \label{emf2drov}\\
\Bigl(\frac{\dot A_3}{A_3}\Bigr)^{\cdot} + \Bigl(\frac{\dot
A_3}{A_3}\Bigr)^2 - 2 \Bigl(\frac{\dot A_3}{A_3}\Bigr)\dot \beta_3 -
\vf_0^2 P_I &=& 0. \label{emf3drov}
\end{eqnarray}
\end{subequations}
In view of \eqref{rela123} from \eqref{emf123drov} we conclude that
\begin{equation}
\dot \beta_1 = \dot \beta_2 = \dot \beta_3, \label{eqdotbets}
\end{equation}
which is equivalent to $b_1 = b_2 = b_3 = b$ in \eqref{betas}, i.e.
\begin{equation}
\beta_1 = \beta_2 = \beta_3 = b t. \label{betasnew}
\end{equation}
As one sees from \eqref{betasnew}  we have isotropy at any given
time.

Now taking into account that both $A_2$ and $A_3$ can be expressed
in term of $A_1$ one could solve only one of the three equations of
\eqref{emf123}. In view of \eqref{rela123new} and \eqref{betasnew}
let us first rewrite \eqref{eminv} as follows
\begin{equation}
I =  - Q A^2 e^{-2bt}, \quad Q = [1 + C_{21}^2  + C_{31}^2 ]. \label{eminvnew}\\
\end{equation}

In view of \eqref{rela123} and \eqref{betasnew} the equation for $A$
now reads
\begin{equation}
A \ddot A + \dot A^2 - 2 b A \dot A - \frac{\lambda}{\vk}\vf_0^2  I
\left[\frac{2}{P} P_I^2 - P_{II}\right] A^2 = 0. \label{eqA}
\end{equation}
In case of $\lambda = 0$ or ${\cG}_{II} =  \left[\frac{2}{P} P_I^2 -
P_{II}\right] = 0$, the equation \eqref{eqA} can be easily solved
\cite{SahaCEJP2011}. In this case we have
\begin{equation}
A \ddot A + \dot A^2 - 2 b A \dot A  = 0. \label{eqA0}
\end{equation}
with the solution
\begin{equation}
A = \sqrt{D^2 e^{2bt} - C/b}, \label{Agen}
\end{equation}
with $D$ and $C$ being the constants of integration. Inserting $A$
from \eqref{Agen} into \eqref{rela123} we find
\begin{equation}
\frac{d P}{d I} = -\frac{1}{\vf_0^2} \left(\frac{\dot A}{A}\right)^2
= -\frac{1}{\vf_0^2} \frac{b^2 Q^2 D^4}{I^2}. \label{Pgen0}
\end{equation}
The second equality was obtained using \eqref{eminvnew}. From
\eqref{Pgen0} one finds that
\begin{equation}
P = \frac{b^2 Q^2 D^4}{\vf_0^2  I} + C_1, \label{Pgen1}
\end{equation}
with $C_1$ being some integration constant. Note that the above
result is valid for both $\lambda = 0$ or ${\cG}_{II} =
\left[\frac{2}{P} P_I^2 - P_{II}\right] = 0$. In case of $\lambda =
0$ the equation \eqref{00new1} takes the form
\begin{equation}
\ddot \alpha - \dot \alpha^2 + \dot \beta_1^2 + \dot \beta_2^2 +
\dot \beta_3^2  =  \vk \vf_0^2 [P + I P_I],\label{00new1a}
\end{equation}
Now taking into account that $\alpha = \beta_1 + \beta_2 + \beta_3 =
3 \beta = 3 b t$, from \eqref{00new1a} we find
\begin{equation}
6 b^2 = - \vk \vf_0^2 C_1. \label{bc}
\end{equation}
That is in case of Einstein theory we get the foregoing solution.
Now let us see, what happens if we consider $f(R,T)$ theory. In this
case the solution \eqref{Agen} occurs due to
\begin{equation}
\frac{2}{P} P_I^2 - P_{II} = 0, \label{Pgen2}
\end{equation}
which leads to
\begin{equation}
P =  \frac{1}{C_2 I + C_3},
\label{Pgen3}
\end{equation}
where $C_2$ and $C_3$ are integration constants. Comparing
\eqref{Pgen1} and \eqref{Pgen3} we conclude that $C_1 = C_3 = 0$ and
$C_2 = \vf_0^2/b^2 Q^2 D^4$. That is in this case we have $P =
\frac{b^2 Q^2 D^4}{\vf_0^2  I}$. Now inserting $\alpha$, $\beta_i$'s
and $P$ into \eqref{00new1a} we find
\begin{equation}
6 b^2 = 0. \label{bc1}
\end{equation}
It mean in this case the spacetime becomes flat. Hence we see that
the consideration of $f(R,T)= R + \lambda f(T)$ leads to the flat
spacetime.

Let us consider the case when $P = I^n$. In this case on account of
\eqref{eminvnew} equation \eqref{eqA} can be written as
\begin{equation}
A \ddot A + \dot A^2 - 2 b A \dot A +
\frac{\lambda}{\vk}\frac{\vf_0^2}{Q} n(n+1)I^n e^{2 bt} = 0.
\label{eqAn}
\end{equation}
On account of \eqref{eminvnew} the foregoing equation can be
rewritten as
\begin{equation}
\ddot X + p \dot X + q X^{n} e^{2 (1 - n) bt} = 0, \label{eqX}
\end{equation}
where we denote $X = A^2$, $ p = - 2b$ and $q = (-1)^n
 \frac{2\lambda}{\vk} \vf_0^2 n (n+1) Q^{n - 1}$. In case of $ n =
1$ from \eqref{eqX} we find
\begin{equation}
\ddot X +  p \dot X + q X = 0, \label{eqXn-1}
\end{equation}
which is the equation for free oscillation. Depending on $p$ and $q$
\eqref{eqXn-1} allows following three solutions:

\begin{equation}\label{X-sol}
X =\left \{
\begin{array}{ll}
C_1 e^{-\frac{D - p}{2} t} + C_2 e^{-\frac{D + p}{2} t} & \quad D^2  > 0 \\
e^{-\frac{pt}{2}}\left(C_1 \cos{\frac{Dt}{2}} + C_2
\sin{\frac{Dt}{2}}\right) & \quad D^2 < 0 \\
e^{-\frac{pt}{2}}\left(C_1 t + C_2\right)  & \quad D^2 = 0,
\end{array}
\right.
\end{equation}
where we denote $D^2 = p^2 - 4q.$ On account of $q = - 4\lambda
\vf_0^2/\vk$ in this case we have $D^2 = 4\left(b^2 + 4\lambda
\vf_0^2/\vk\right).$

On the other hand inserting $P = I^n$  into \eqref{00new1} we find
\begin{equation}
6 b^2 = \vf_0^2 (1 + n) \left[\lambda (1 + n) - \vk\right] I^n.
\label{bpn}
\end{equation}
Since $b$ is a constant, this \eqref{bpn} can be valid only if $n =
0$ which gives $6 b^2 = \vf_0^2 \left[\lambda - \vk\right]$. It
means in the case considered we have $P = 1 \Rightarrow {\cG} = 1$,
i.e. we come to the case of minimal coupling.

\section{conclusion}

Within the scope of $f(R,\,T)$ theory of gravity an interacting
system of scalar and electromagnetic fields has been thoroughly
investigated. As one can expect, the case with $f(R,T) = R + \lambda
f(T)$ leaves the system qualitatively same though at places bring
some additional restrictions on the solutions.

\acknowledgements{This work is supported in part by a joint
Romanian-LIT, JINR, Dubna Research Project, theme no.
05-6-1119-2014/2016. Taking the opportunity I would also like to
thank the reviewers for some helpful discussions and references.}


\begin{thebibliography}{99}

\bibitem{Perl1}
S. Perlmutter {\it et al.} (Supernova Cosmology Project
Collaboration), Nature {\bf 391}, 51 (1998).
\bibitem{Perl2}
S. Perlmutter {\it et al.} (Supernova Cosmology Project
Collaboration), Astrophys. J. {\bf 517}, 565 (1999).
\bibitem{Perl3}
S. Perlmutter {\it et al.}, Astrophys. J. {\bf 598}, 102 (2003).
\bibitem{Reiss1}
A.G. Riess {\it et al.} (Supernova Search Team Collaboration),
Astron. J. {\bf 116}, 1009 (1998).
\bibitem{Reiss2}
A.G. Riess {\it et al.} (Supernova Search Team Collaboration),
Astrophys. J. {\bf 607}, 665 (2004).
\bibitem{Tonry}
J.L. Tonry {\it et al.} (Supernova Search Team Collaboration),
Astrophys. J. {\bf 594}, 1 (2003).
\bibitem{Clocch}
A. Clocchiatti {\it et al.} (High Z SN Search Collaboration),
Astrophys. J. {\bf 642}, 1 (2006).

\bibitem{carrol} S.M. Carrol, V. Duvvuri, M. Troden, and S.M. Turner
{\it Pjys. Rev. D} {\bf 70}, 043528 (2004).

\bibitem{harko} T. Harko, F.S.N. Lobo, S. Nojiri, and S.D. Odintsov {\it Phys. Rev. D}, {\bf 84},
024020 (2011)]

\bibitem{SahaCEJP2011} Rybakov Yu.P., Shikin G.N., Popov Yu. A., Saha B. {\it Cent.
Eur. J. Phys.} {\bf 9} 1165 (2011)

\bibitem{mie1} Mie G. {\it Annalen der Physik}, {\bf 37}, 511 (1912)

\bibitem{mie2}  Mie G. {\it Annalen der Physik}, {\bf 39}, 1 (1912)

\bibitem{SahaIJTP1997} Rybakov Yu.P., Shikin G.N., Saha B. {\it Int.
J. Theor. Phys.} {\bf 36} 1475 (1997)





\end{thebibliography}
\end{document}